\documentclass[12pt]{article}
\usepackage{graphicx,amsfonts}
\usepackage{longtable}
\usepackage{amsmath}
\usepackage[all]{xy}

\newcommand{\be}{\begin{equation}}
\newcommand{\ee}{\end{equation}}
\newcommand{\ba}{\begin{eqnarray}}
\newcommand{\ea}{\end{eqnarray}}
\newcommand{\baa}{\begin{eqnarray*}}
\newcommand{\eaa}{\end{eqnarray*}}

\begin{document}

\title{The tunneling conductance between a superconducting STM tip and an out-of-equilibrium carbon nanotube}
\author{Cristina Bena$^{a,b}$\\
{\small \it $^a$Laboratoire de Physique des Solides, Universit\'e Paris-Sud},
\vspace{-.1in}\\{\small \it  B\^at.~510, 91405 Orsay, France}\\
{\small \it $^b$Institut de Physique Th\'eorique, CEA/Saclay, CNRS, URA 2306},
\vspace{-.1in}\\{\small \it  Orme des Merisiers, F-91191 Gif-sur-Yvette, France}}
\maketitle

\begin{abstract}
  We calculate the current and differential conductance for the
  junction between a superconducting (SC) STM tip and a Luttinger
  liquid (LL).  For an infinite single-channel LL, the SC coherence peaks are
  preserved in the tunneling conductance for interactions weaker than
  a critical value, while for strong interactions ($g <0.38$), they
  disappear and are replaced by cusp-like features. For a finite-size
  wire in contact with non-interacting leads, we find however that the
  peaks are restored even for extremely strong interactions. In the
  presence of a source-drain voltage the peaks/cusps split, and the
  split is equal to the voltage. At zero temperature, even very strong
  interactions do not smear the two peaks into a broader one; this
  implies that the recent experiments of Y.-F. Chen et. al. (Phys.
  Rev. Lett. {\bf 102}, 036804 (2009)) do not rule out the existence
  of strong interactions in carbon nanotubes.
\end{abstract}

\maketitle

Scanning tunneling spectroscopy (STM) is becoming an important tool
for accessing the electronic properties of low-dimensional systems.
Thus, in the past, the local density of states (DOS) for various
materials such as the cuprates \cite{sc}, graphene \cite{graphene},
and carbon nanotubes\cite{nt} has been studied, both theoretically and
experimentally. In general the STM tip is assumed to be a
non-interacting metal with a constant DOS, and the system to be
analyzed is in equilibrium (the voltage is constant throughout the
sample). This allows one to extract the unknown DOS directly from the
STM tunneling conductance.

Recently a few studies have also concentrated on tips that are not
normal metals, and that may have an intrinsic variation in the DOS,
such as superconducting (SC) STM tips \cite{pothier,mason}. These
experiments have the potential of measuring not only the DOS, but also
other quantities such as the Fermi distribution. The first such experiment \cite{pothier}
looked at a non-interacting system, and found
that in equilibrium the differential conductance from a SC STM tip
shows the characteristic SC gap and coherence peaks, while in the
presence of a voltage bias each peak splits into two peaks, with the
distance between them being given by the applied voltage. For a
non-interacting system this splitting can be traced back to an
out-of-equilibrium double-step Fermi distribution.

Similar experiments have been performed recently on carbon nanotubes
\cite{mason}, but the interpretation of these measurements is not so
straightforward, due to the presence of strong electronic interactions
which make the elementary excitations no longer fermionic but rather
fractionally-charged. One might naively expect that the Fermi
distribution is ill-defined in these systems, and that the strong
electronic interactions contribute to the smearing of the two SC
coherence peaks into a single one. On the other hand the experiments
of \cite{mason} show the presence of two coherence peaks, which, when
combined with this naive expectation, appear to imply that the
interactions in nanotubes are weak !

To fully assess the implications of this experiment, and to test
whether the expectation that the peaks merge is borne out of rigorous
calculations, one needs to study theoretically the injection of
electrons into a strongly-interacting out-of-equilibrium
one-dimensional system. This is rather a challenging problem, that has
not begun to be addressed theoretically until quite recently
\cite{stmold,dolcini,gefen,ines}.

We study the injection from a SC tip into an out-of-equilibrium LL,
and we show that, even in the presence of very strong interactions,
the tunneling conductance is very similar to that of a non-interacting
wire. Our calculations therefore demonstrate that the apparently
non-interacting features observed in \cite{mason} do not rule out the
existence of strong interactions in carbon nanotubes.

We begin by analyzing an infinite single-channel LL wire in equilibrium, and we find
that the tunneling conductance displays SC coherence peaks for
interactions weaker than a critical value, corresponding to a
fractional charge parameter, $g$, larger than $g_c=0.38$. Beyond this
value the coherence peaks disappear, and are replaced by cusp-like
features. Upon taking into account the finite size of the wire \cite{safi,maslov,finitesize}
and the presence of non-interacting contacts (as in the experiment of
\cite{mason}) we find that the peaks reappear for any value of $g$.
Furthermore, the tunneling conductance shows also oscillations with a
period inverse proportional to the length of the wire and has a
background power-law dependence whose exponent depends on the strength
of the interactions.

If a voltage is applied between the ends of the wire, the peak/cusp
features split. The magnitude of the split is exactly equal to the
applied voltage. At zero temperature we never see a smearing of the
two peaks into a single one, regardless of the strength of the
interaction.  Hence, the expectation that interactions cause the two
peaks to merge into one is not correct\footnote{This is consistent with the theoretical calculations in Ref.~\cite{gefen} showing a double-step Fermi distribution in an out-of-equilibrium LL.}.
This in turn implies that the
presence of two peaks found in \cite{mason} can not be used to rule
out the presence of strong interactions in nanotubes !

Our calculations show that the experimental setup of \cite{mason} can be used
to asses the strength of the interactions if one focuses instead on the
power-law dependence of the density of states. If one measured the
density of states for a wide-enough range of voltages (up to tens of meV)
the data should allow to extract the value of the interaction parameter $g$.

\vspace{.3in}

A quantum wire connected to metallic leads is described by the Hamiltonian
\begin{equation}
{\mathcal{H}} ={\mathcal{H}}_{0}  \, +
\, {\mathcal{H}}_{V} \; . \label{L}
\end{equation}
Here  ${\mathcal{H}}_{0}$ describes the interacting wire and
the leads in the framework of the inhomogeneous single-channel LL model \cite{safi,maslov,finitesize,dolcini} (we will discuss at the end how our results are affected by the presence of the extra channels of conduction in nanotubes).
Explicitly:
\begin{eqnarray}
{\mathcal{H}}_0 &=&\frac{\hbar v_F}{2}  \int_{-\infty}^{\infty}
 dx \left[ \Pi^2 + \frac{1}{g^2(x)}
(\partial _x\Phi )^2\right] \\
{\mathcal{H}}_{V}  &=&    - \int_{-\infty}^{\infty}
\frac{dx}{\sqrt{\pi}} \,\mu(x) \,
\partial_x \Phi(x,t) \; , \label{LV}
\end{eqnarray}
where ${\mathcal{H}}_{V}$ describes the chemical potential applied to the wire.
The interaction parameter $g(x)$ is space-dependent and its value is
$g$ in the bulk of the wire, and 1 in the leads. For convenience, the end-points
of the wire are denoted by $x_1=-L/2$ and $x_2=L/2$. The chemical
potential is chosen such that $\mu(x)=-V_{SD}/2$ for $x<-L/2$ (left
lead), $\mu(x)=V_{SD}/2$ for $x>L/2$ (right lead), and $\mu(x)=0$ for
$-L/2<x<L/2$

Tunneling is allowed between the wire and a superconducting tip at at
$x=0$.  A schematic view of the system is shown in Fig.~\ref{setup}.
The voltage of the tip is fixed at $V$.

\begin{figure}[htbp]
\vspace{0.3cm}
\begin{center}
\includegraphics[width=10cm]{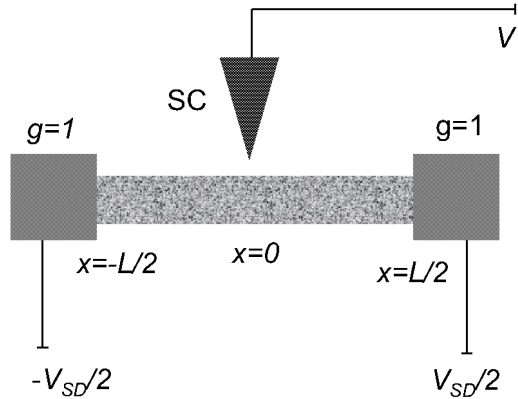}
\vspace{-0.15in} \caption{\small  A quantum wire adiabatically coupled to metallic leads. The leads are held at different chemical potentials $\mu_1=-eV_{SD}/2$ and $\mu_2=eV_{SD}/2$. The electrons can tunnel from and into a superconducting STM tip at $x=0$. The voltage of the tip is fixed at $V$.}
\label{setup}
\end{center}
\end{figure}

The Hamiltonian for the SC tip is assumed to be of the BCS type with a linear dispersion ($\epsilon_k\propto k$) and with a SC BCS coupling,
\be
H_{tip}=\sum_{k} \sum_{\alpha=\uparrow,\downarrow} \epsilon_k c^\dagger_{k\alpha} c_{k\alpha}+(\Delta c_{k \uparrow} c_{-k \downarrow} +h.c.).
\ee
Our results do not depend on the details of the dispersion and of the dimensionality of the tip, as long as the
 tip DOS has the standard BCS form $\nu_t(\epsilon)=\theta(\epsilon-\Delta) \epsilon/\sqrt{\epsilon^2-\Delta^2}$.

The tunneling between the tip and the wire is described by the local Hamiltonian
\be
H_{T}=t c^{\dagger}_{\alpha}(x=0)[\Psi^R_{\alpha}(x=0)+\Psi^L_{\alpha}(x=0)]+h.c
\ee
where $\Psi^{R/L}_{\alpha}(x)$ are the chiral fermionic interacting operators that can be related to the free bosonic modes of the LL via bosonization \cite{fisher}.

We use the non-equilibrium Keldysh formalism and we perform a
perturbative expansion up to second order in the tunneling coefficient
$t$, using the formalism developed in Ref.~\cite{dolcini} to calculate
the tunneling current between a normal tip and an out-of-equilibrium LL.
We find that the tunneling current between the SC
tip and the wire at zero temperature is

\be
I \propto |t|^2 \int_0^{\infty} dt Q_T(t) {\rm Im} \{\exp[{\cal R}(t)+i {\cal I}(t)] {\cal F}_s(t)\}\, ,
\ee
where for a finite-size wire \cite{dolcini}
\be
Q_{T}(t) = 2 \cos \left( \frac{V_{SD} t}{2} \right)\sin(V t)\,,
\ee
and
\ba
\mathcal{R}(t)
&=& -\frac{1}{32 \pi} \left\{ \left(g+g^{-1}-2\right)
\sum_{m \in Z_{\rm even}} \rho^{|m|} \ln{ \frac{\alpha_{W}^2+(\tau + m)^2}{\alpha_{W}^2+m^2} }  \right. \\
& & +\left(g+g^{-1}+2 \right)
\sum_{m \in Z_{\rm even}} \rho^{|m|}\ln{ \frac{\alpha_{W}^2+(\tau - m)^2}{\alpha_{W}^2+m^2} } \\
& & + \left(g-g^{-1}\right) \sum_{m \in
Z_{\rm odd}} \rho^{|m|} \left( \ln{
\frac{\alpha_{W}^2+(\tau + m)^2}{\alpha_{W}^2+
m^2} } + \ln{ \frac{\alpha_{W}^2+(\tau -
m)^2}{\alpha_{W}^2 + m^2} }\right)\}
\ea

\ba
\mathcal{I}(t)&=&
-\frac{1}{16 \pi} \left\{ \left(g+g^{-1}-2 \right)
\sum_{m \in Z_{\rm even}} \rho^{|m|} \arctan{ \left( \frac{\tau+ m}{\alpha_{W}} \right) }  \right. \\
& & + \left(g+g^{-1}+2 \right)
\sum_{m \in Z_{\rm even}} \rho^{|m|} \arctan{ \left( \frac{\tau - m}{\alpha_{W}} \right) } \\
& & \left. +\left(g-g^{-1}\right)
\sum_{m \in Z_{\rm odd}} \rho^{|m|} \left[ \arctan{ \left(
\frac{\tau + m}{\alpha_{W}}\right) }  + \arctan{
\left(\frac{\tau - m}{\alpha_{W}}\right)} \right] \right\} \,,
\ea
with $\tau \equiv t v_F/g L$, $\rho \equiv (1-g)/(1+g)$, and $\alpha_W$ is a dimensionless short-length cutoff. The quantity $Q_T(t)$ incorporates the effects of the applied tip voltage and source-drain voltage, while $\mathcal{R}(t)$ and $\mathcal{I}(t)$ are related to the Green's functions of the system and incorporate information about the finite size via $\tau$. For an infinite wire one should substitute these quantities by:
\begin{equation}
\mathcal{R}(t)= -\frac{1}{16 \pi}  \left(g+g^{-1}\right)
\ln{ \frac{a_{W}^2+t^2}{a_{W}^2} }
\end{equation}

\begin{equation}
\mathcal{I}(t)=
-\frac{1}{8 \pi}  \left(g+g^{-1}-2 \right)
\arctan{ \left( \frac{t}{a_{W}} \right) }\, .
\end{equation}
Here $a_{W}$ is a short-time (high-energy) cutoff for the infinite wire.

For the superconducting tip, by taking a Fourier transform of the usual SC Green's function we obtain
\be
{\cal F}_{s}(t)=\Delta K_1[(i t + a_T) \Delta]\,,
\ee
where $a_T$  is a short-time (high-energy) cutoff for the tip, and $K_1$ is the first Bessel $K$ function.

For a finite wire we can use the formulae above to compute the
differential conductance $dI/dV$ numerically. For an infinite wire
this calculation can be performed analytically, by rewriting
(along the lines of \cite{gefen}) the derivative of the tunneling current as:
\ba
\frac{d I}{d V}\propto  \sum_{\eta=\pm 1}|t|^2 \int_{0}^{\infty} d\epsilon \frac{d \nu_{t}(\epsilon)}{d \epsilon}(|\epsilon+V+\eta V_{SD}/2|^{\gamma}-|\epsilon-V-\eta V_{SD}/2|^{\gamma})
\ea
where $\gamma \equiv (g+1/g-2)/2$ and $\nu_t(\epsilon)$ is the density of states of the superconducting tip.
This integral now gives:
\ba
\frac{d I}{d V}\propto  && i \sum_{\eta=\pm 1}|t|^2[|\Delta - V-\eta V_{SD}/2|^{\gamma - 1/2} {_2F_1}(3/2, -1/2; 3/2 -\gamma; \Delta/2 + V/2+\eta V_{SD}/4) + \nonumber\\
&&|\Delta + V+\eta V_{SD}/2|^{\gamma - 1/2} {_2F_1}(3/2, -1/2; 3/2 - \gamma; \Delta/2 - V/2-\eta V_{SD}/2)]
\label{inf}
\ea
where $_2F_1$ is the corresponding hypergeometric function.

We expand this function (in equilibrium -- $V_{SD}=0$) close to the
singularity at $V=\Delta$ and at $V\gg\Delta$. When
$V\approx \Delta$:
\be
\frac{d I}{d V}\propto  (V-\Delta)^{\gamma-1/2}\,.
\label{powerlawdecay}
\ee
For non-interacting systems $\gamma=0$, and the tunneling
conductance diverges as expected: $\frac{d I}{d V} \propto (V-\Delta)^{-1/2}$. In
the presence of interactions this exponent is modified. For very strong
interactions $\gamma$ increases, the exponent changes sign, and the divergence is replaced
by a power-law-type cusp. This regime is achieved when $\gamma >1/2$,
which corresponds to $g<0.38$.

When $V\gg\Delta$ we obtain
\be
\frac{d I}{d V}\propto  V^{\gamma}\,.
\ee
For non-interacting systems the tunneling conductance is constant at
large V. In the presence of interactions the conductance acquires a
power-law background with a positive exponent $\gamma$ as expected for
tunneling into a LL.

In Figs.~2 a) and b) we plot the tunneling conductance for $V_{SD}=0$,
for $g=1$ and $g=0.5$. We take $\Delta=1$ in arbitrary units. In
Figs.~2 c) and d) we plot the corresponding tunneling
conductance\footnote{The source-drain voltage is applied such that the
  average of the right and left potentials is zero, and the bulk of
  the wire remains neutral. If the voltage of the bulk of the wire
  becomes non-zero (because of, say, a gate voltage), the curves
  will shift along the voltage axis such that the symmetry point of the curve will no longer be at zero tip voltage, but at a value of the tip voltage equal to the gate voltage.} when $V_{SD}=\Delta$. As described
above, for $g=0.5>g_c$ the SC coherence peaks are preserved in the
spectrum. When the wire is out of equilibrium, Eq.~(\ref{inf}) implies that
the tunneling conductance is given by the
sum of two tunneling conductances, which are shifted respectively by $\pm V_{SD}/2$.

\begin{figure}[htbp]
\vspace{0.3cm}
\begin{center}
\includegraphics[width=12cm]{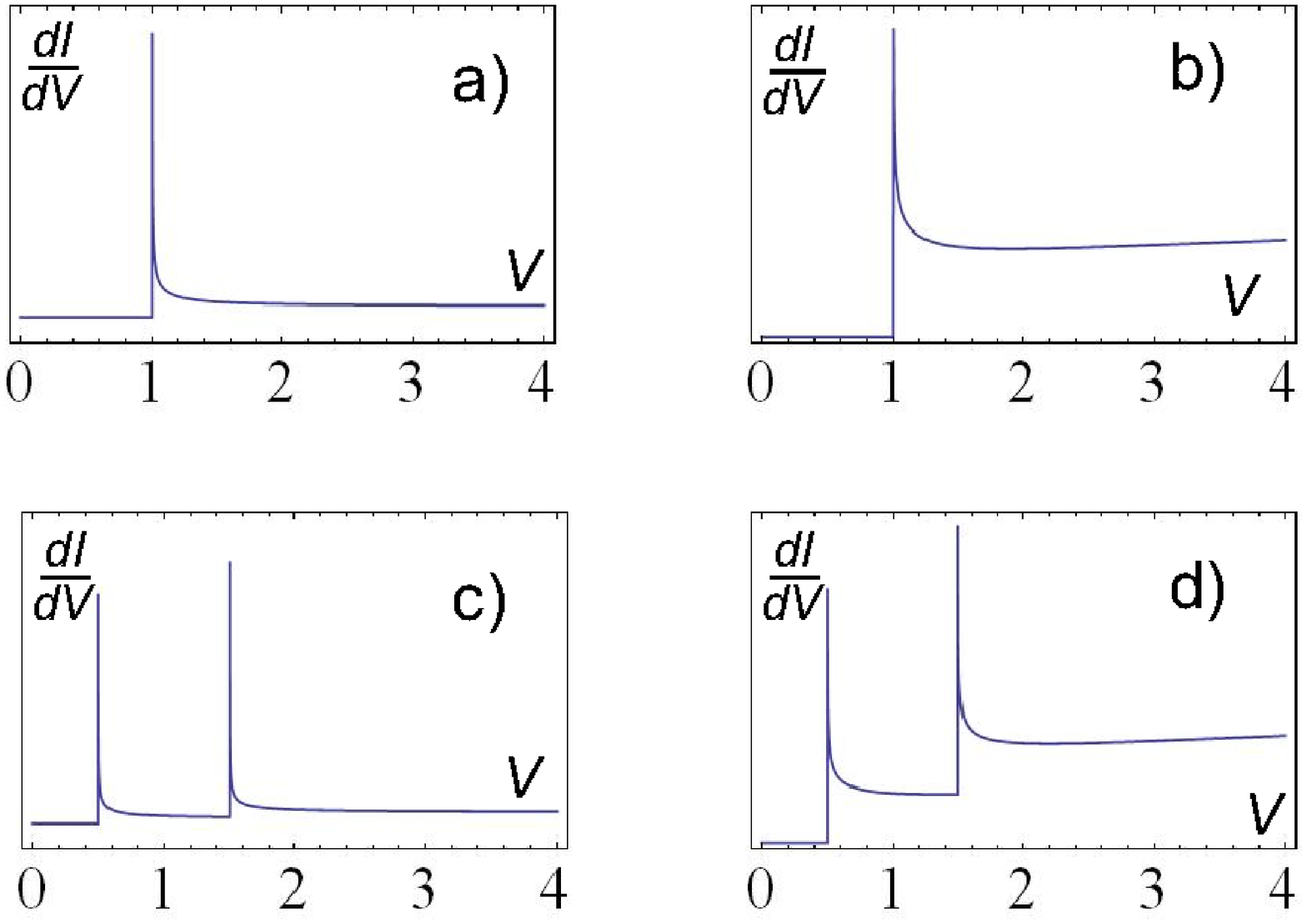}
\vspace{-0.15in} \caption{\small The tunneling conductance  from an STM tip into an infinite Luttinger liquid (in arbitrary units) for LL interaction parameters $g=1$ (in a) and c)) and $g=0.5$ (in b) and d)). The source-drain voltage $V_{SD}$ is zero for a) and  b), and is $V_{SD}=\Delta=1$ for c) and d).}
\label{setup}
\end{center}
\end{figure}

In Figs.~3 a) and c) we repeat the same calculation for $g=0.3<g_c$.
We see that indeed the peaks disappear and are replaced by cusp-like features, corresponding to a
power-law decay of the tunneling conductance (\ref{powerlawdecay}).

Our results change significantly if we take into account the finite
size of the wire and the presence of metallic contacts. The integrals
can now only be done numerically. In Fig.~3 b) and d) we plot the
tunneling conductance for $g=0.3$ for a finite-size wire with
characteristic finite-size energy scale $\omega_L=\hbar v_F/g L=
\Delta/3$. We see that the peaks that disappear at infinite lengths
are restored for a finite-size system. This is consistent with
previous observations for nanotubes \cite{safi,maslov,finitesize}: when
$V-\Delta\ll\omega_L$ the physics is dominated by long wavelengths and is
hence non-interacting.

Most importantly, for both infinite and finite systems, the two
features at $\Delta\pm V_{SD}/2$ are not smeared into a single one,
regardless of the strength of the interactions. Thus, as far
as the peaks are concerned, even a very strongly interacting wire
behaves as non-interacting. This is consistent with
previous observations of a double-step out-of-equilibrium Fermi distribution \cite{gefen},
and explains why the features
observed in Ref.\cite{mason} appear for both interacting and
non-interacting systems. The only signatures of interactions
are the disappearance of the peaks at very small $g$ for the infinite
wire, and the power-law dependance of the tunneling conductance.

We should note that for a nanotube the parameter $\gamma$ is reduced
from $(g+1/g-2)/2$ to $(g+1/g-2)/8$ because of the extra three
channels of conduction \cite{bf}. Hence, the critical value of $g$
below which the peaks disappear is $g_{\rm cn}\approx 0.17$.
Nevertheless, the finite-size effects are always relevant for
non-chiral LLs, and hence we expect that the spectra presented in Figs.~3 b) and d), characterized by oscillations and SC
coherence peaks, should describe the tunneling conductance between a SC tip and a nanotube in all situations.

\begin{figure}[htbp]
\vspace{0.3cm}
\begin{center}
\includegraphics[width=12cm]{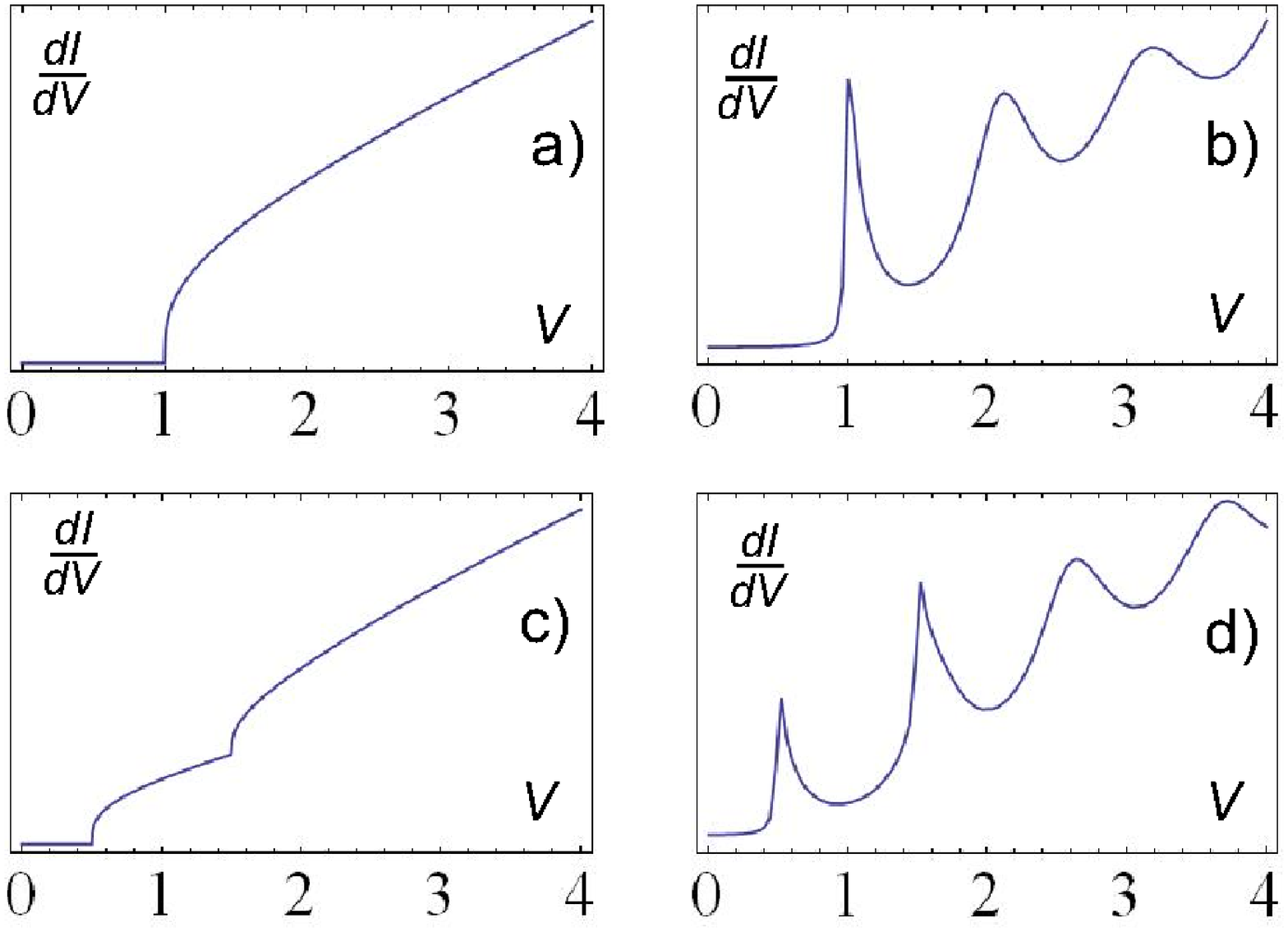}
\vspace{-0.15in} \caption{\small The tunneling conductance from a SC STM tip into an infinite (a) and c)) and finite-sized (b) and d))  $g=0.3$ Luttinger liquid. The source-drain potential $V_{SD}$ is zero for a) and b) and is $V_{SD}=\Delta=1$ for c) and d). The energy corresponding to the finite size of the wire, $\hbar v_F/g L$, is taken to be $\Delta/3$.}
\label{setup}
\end{center}
\end{figure}

For a physical set of parameters, $\Delta=V=1.5meV$ and $\hbar v_F/g L
\approx 0.5 meV$ (corresponding to a tube length of $\approx 4 \mu m$
at $g=0.25$), our results for the tunneling conductance reproduce very well those measured in \cite{mason} (compare for example Figs.~3 b) and 3 d) here with Figs.~2 c) and 3 in \cite{mason}).

To summarize, we have calculated the dependence on voltage of the STM
tunneling conductance between a superconducting STM tip and an
out-of-equilibrium Luttinger liquid. We have found that for an
infinite single-channel LL in equilibrium, one can observe SC
coherence peaks for interactions weaker than a critical value ($g <
0.38$). Beyond this value, the coherence peaks disappear and are
replaced by cusp-like features. We have also found that the peaks are
restored if one takes into account the finite size of the LL and the
presence of non-interacting contacts. For both infinite and
finite-sized wires, the tunneling conductance shows a background
power-law dependence, with an exponent that can be used to determine
the strength of the interactions.

In the presence of an applied voltage between the ends of the wire,
the tunneling-conductance features (peaks or cusps) split in two, and
the magnitude of the split is equal to the voltage. Thus, there is no
smearing of two peaks into a single one, regardless of the strength of
interactions.  This implies that the measurements of Ref.\cite{mason}, that reveal
features which appear naively to be non-interacting, do not in fact
rule out the presence of strong interactions in carbon nanotubes.

{\bf Acknowledgments} We would like to thank N. Mason, C. H. L. Quay, J. -D. Pillet, I. Safi and S. Vishveshwara for helpful discussions.

\end{document}